\documentclass[12pt]{article}
\usepackage[utf8]{inputenc}
\usepackage{graphicx}
\graphicspath{{figures/}}
\usepackage{color}
\usepackage{amsmath}
\usepackage{float}
\usepackage{hyperref}
\usepackage{comment}
\usepackage[margin=1in]{geometry}
\usepackage[style=nature,doi=false,isbn=false,url=false]{biblatex}
\usepackage[labelfont=bf]{caption}

\usepackage[utf8]{inputenc}
\usepackage[T1]{fontenc}
\usepackage{mathptmx}
\usepackage{multirow}
\usepackage{chemformula}
\usepackage{upgreek}
\usepackage{float}
\usepackage{subcaption}
\usepackage{cleveref}
\usepackage[normalem]{ulem}
\useunder{\uline}{\ul}{}
\usepackage[labelfont=bf, format=plain]{caption}

\title{{\bf Compound Mutations in the Abl1 Kinase Cause Inhibitor Resistance by Shifting DFG Flip Mechanisms and Relative State Populations
}}

\begin{document}
\maketitle


\begin{center}
\noindent
\author{{\bf \large Gabriel Monteiro da Silva } \\
\textit{Department of Molecular Biology, Cell Biology, and Biochemistry \\ Brown University, Providence, RI 02912, USA}
\vspace{5mm} \\

{\bf \large Kyle Lam} \\
\textit{Department of Chemistry} \\
\textit{Brown University, Providence, RI 02912, USA}
\vspace{5mm} \\
{\bf \large David C. Dalgarno} \\
\textit{Dalgarno Scientific LLC} \\ \textit{Brookline, MA 02446, USA} \vspace{5mm} \\
{\bf \large Brenda M. Rubenstein*}\\ 
\textit{Department of Chemistry \\ Department of Molecular Biology, Cell Biology, and Biochemistry} \\ 
\textit{Brown University, Providence, RI 02912, USA} \\ 
}
\end{center}
\section{Abstract}
The intrinsic dynamics of most proteins are central to their function. Protein tyrosine kinases such as Abl1 undergo significant conformational changes that modulate their activity in response to different stimuli. These conformational changes constitute a conserved mechanism for self-regulation that dramatically impacts kinases' affinities for inhibitors. Few studies have attempted to extensively sample the pathways and elucidate the mechanisms that underlie kinase inactivation. In large part, this is a consequence of the steep energy barriers associated with many kinase conformational changes, which present a significant obstacle for computational studies using traditional simulation methods. Seeking to bridge this knowledge gap, we present a thorough analysis of the ``DFG flip'' inactivation pathway in Abl1 kinase. By leveraging the power of the Weighted Ensemble methodology, which accelerates sampling without the use of biasing forces, we have comprehensively simulated DFG flip events in Abl1 and its inhibitor-resistant variants, revealing a rugged landscape punctuated by potentially druggable intermediate states. Through our strategy, we successfully simulated dozens of uncorrelated DFG flip events distributed along two principal pathways, identified the molecular mechanisms that govern them, and measured their relative probabilities. Further, we show that the compound Glu255Lys/Val Thr315Ile Abl1 variants owe their inhibitor resistance phenotype to an increase in the free energy barrier associated with completing the DFG flip. This barrier stabilizes Abl1 variants in conformations that can lead to loss of binding for Type-II inhibitors such as Imatinib or Ponatinib. Finally, we contrast our Abl1 observations with the relative state distributions and propensity for undergoing a DFG flip of evolutionarily-related protein tyrosine kinases with diverging Type-II inhibitor binding affinities. Altogether, we expect that our work will be of significant importance for protein tyrosine kinase inhibitor discovery, while also furthering our understanding of how enzymes self-regulate through highly-conserved molecular switches.


\section{Introduction \label{intro}}

Protein kinases (PKs) are the master regulators of eukaryotic life \cite{Manning2002}. Through their phosphotransfer activity, kinases can induce or suppress a wide range of cellular processes, from proliferation to apoptosis \cite{Cipak2022}. Unsurprisingly, given their central role in organisms, the activity of PKs must be tightly regulated to maintain or achieve homeostasis and their dysregulation often leads to severe disease \cite{Cipak2022, Greenman2007, Pavlovsky2019, Greuber2013, Shah2002}. The role of the Abl1 protein tyrosine kinase in cancers caused by the Philadelphia chromosome phenotype is perhaps the most widely-known example of kinase-mediated oncogenesis \cite{DeBraekeleer2011}. Normally, Abl1 plays a crucial role in various cellular processes, such as cell division and differentiation \cite{Wang2014}. However, oncogenesis manifests when the gene that codes for Abl1 is disturbed such as by the reciprocal translocation observed in the Philadelphia chromosome phenotype, which results from a fusion of the BCR and Abl1 genes \cite{DeBraekeleer2011, Pavlovsky2019, Piedimonte2019}. The protein product that results from this fusion is a constitutively-active form of Abl1, disrupting homeostasis and leading to disease \cite{Kang2016}. Importantly, the presence of the BCR-Abl1 fusion gene is a hallmark of chronic myeloid leukemia and other forms of cancer \cite{Kang2016}.

In many organisms, kinase regulation is achieved through interlocking and often redundant mechanisms \cite{Tong2014, Xie2020}. One example of such a mechanism is the steric blocking of ATP-binding sites, in which kinases such as the Abl1 tyrosine kinase can have their active sites blocked by binding partners or by their own mobile segments, rendering them unable to bind sugar phosphates and inhibiting phosphorylation \cite{Meng2015, Shan2009, Lakkaniga2019}. Abl1 inactivation due to its own mobile segment shifting to block its own active site is a textbook example of inhibition by intrinsic dynamics \cite{Shukla2023}.

Broadly, Abl1 and other kinases are capable of self-inhibition via intrinsic dynamics through different processes, perhaps the most well-studied of which is the conformational change known as the DFG flip \cite{Meng2015, Shan2009, Lakkaniga2019}. Aptly named, the DFG flip is characterized by significant torsions in the backbone of residues 381-388 in Abl1 \cite{Meng2015}, leading to a complete inversion of the orientations of the side chain atoms of Asp381 (D) and Phe382 (F) in the plane formed with the side chain of Ala380 (Figure 1), and similar side chain inversion of downstream A-loop residues 383 to 388. When completed, this inversion inserts Phe382 into Abl1's ATP-binding site (previously occupied by the catalytic Asp381), inhibiting the enzyme.

\begin{figure}[H]
\centering
    \includegraphics[width=500pt]{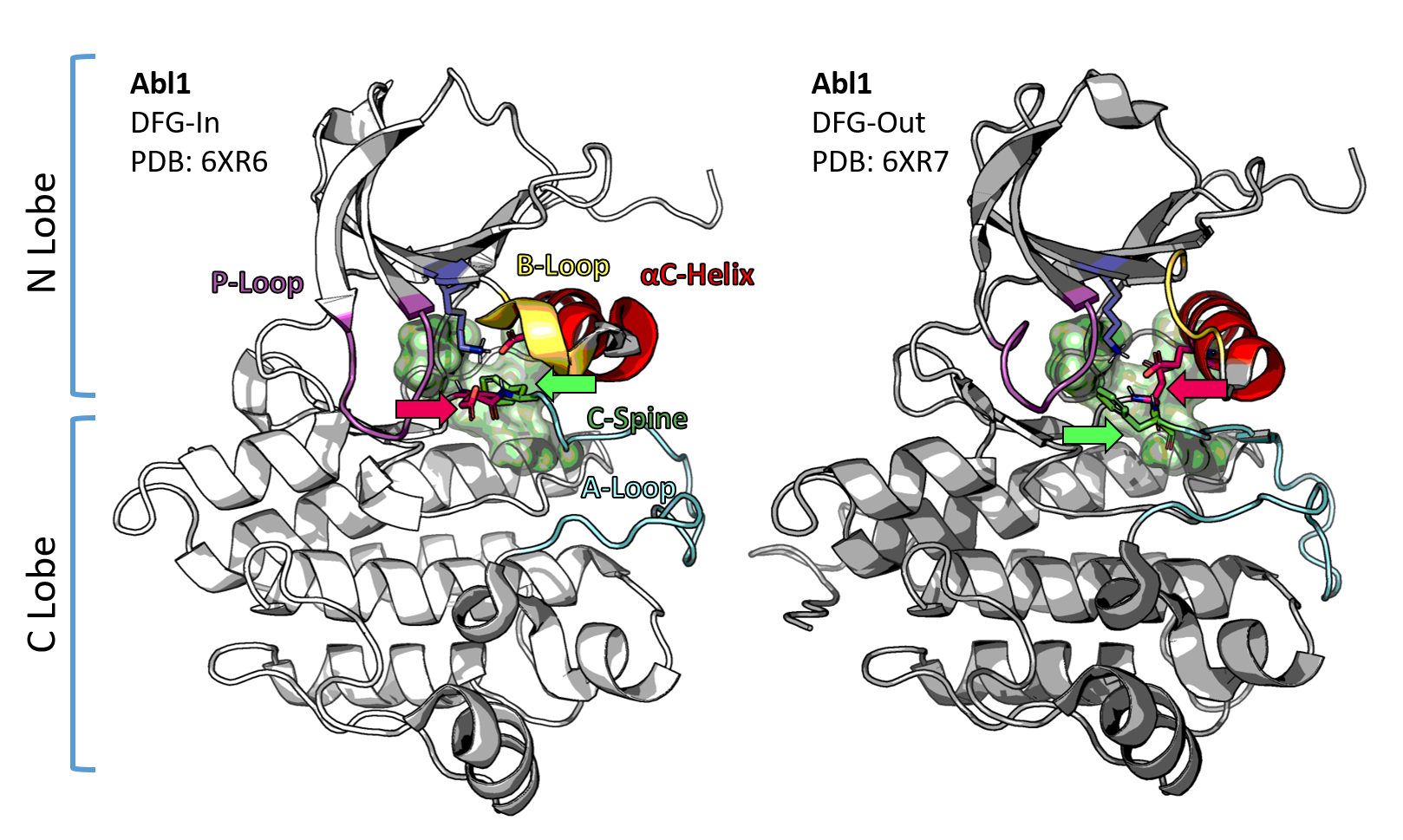}
    \caption{Structural representations of the wild-type Abl1 kinase core in either the DFG-in (left, PDB 6XR6) or DFG-out (right, PDB 6XR7) states as elucidated through nuclear magnetic resonance experiments. For ease of visualization, key residues are represented as sticks (Asp381, red; Phe382 green, both indicated by broad arrows of matching colors; Lys271, dark blue; Glu286, dark red). Additional elements important for kinase function and involved in the DFG flip are colored in pink (P-Loop, residues 249-255), orange (B-Loop, residues 275-278), red (\(\alpha\)C-Helix, residues 280-293), green (hydrophobic C-spine), and cyan (activation loop, residues 385-400).} 
     \label{fig_1}
\end{figure}

Elucidating the structural consequences of this conformational change was crucial for furthering our understanding of how the different states occupied by kinases relate to their affinity for inhibitors, which is of continuing importance for minimizing the broad public health burden of oncogenic kinases \cite{Wilson2015}. This burden is made worse due to the challenges in designing reliable kinase inhibitors due to the high level of structural similarity across the human kinome \cite{Hanson2019, Shah2018, Modi2021}. Since most kinases share a highly conserved catalytic core, strong off-target interactions can be exceedingly common, with undesirable effects \cite{Hanson2019}. This prevalence of off-target effects is what made the discovery of Imatinib, a potent protein kinase inhibitor with extremely high selectivity for Abl1 \cite{Cohen2021}, such an important breakthrough. Imatinib's strong preference for Abl1 puzzled researchers for years, as even structural data of Imatinib-bound complexes did not explain why Imatinib's affinity for Abl1 was significantly higher than for related kinases such as Src \cite{Manley2002}.

This apparent paradox was solved when dynamics were taken into account \cite{Wilson2015}. Specifically, researchers found through nuclear magnetic resonance (NMR) experiments that Abl1 occupied structural conformations that interacted strongly with Imatinib far more frequently than other kinases \cite{Wilson2015}. Specifically, Imatinib binds preferentially to the DFG out (inactive) conformation of kinases, and Abl1 is known to occupy this conformation significantly more frequently than Src or other kinases \cite{Wilson2015}. Additionally, Imatinib induces binding-favorable conformations in Abl1 more readily than in other kinases \cite{Wilson2015, Ayaz2023}.

The mechanism of Imatinib's selectivity represents a cautionary tale on the importance of accounting for conformational dynamics when studying protein systems. While researchers have routinely used computational simulations to explore the binding energies of kinase inhibitors against previously-known kinase conformations, fewer studies have sampled the entire inactivation pathway and other intrinsic conformational changes in Abl1 \cite{Thomas2021}. Although unfortunate given the importance of studying intrinsic kinase dynamics for properly predicting the binding of inhibitors, this research gap is not surprising. Biophysical studies have found that the energy barrier associated with Abl1 inactivation exceeds 30 kcal/mol, representing a free energy barrier that is too steep to be routinely scaled within the timescales accessible with traditional simulation methods such as molecular dynamics \cite{Xie2020}. Accordingly, previous computational studies seeking to understand the DFG flip have relied either on engineered mutations and/or accelerated dynamics methods that use biasing forces to reach the necessary timescales for the flip \cite{Lakkaniga2019, Oruganti2021, Narayan2020, Shan2009}. Although these studies represent major contributions to the field, the application of these methods comes with substantial caveats that limit the mechanistic insights that can be drawn from the resulting dynamics, as they might lead to biasing of pathways, inaccurate reporting of transition state frequencies/energies, or improper sampling due to hysteresis \cite{Bussi2020}.

Despite sampling challenges, fully understanding the details of the Abl1 inactivation pathway is of vital importance for drug discovery \cite{Liu2006}. Atomistic simulations of Abl1 inactivation without biasing forces would provide an unparalleled wealth of information regarding the enzyme, enabling the identification of allosteric networks and potentially druggable intermediate conformations, generalization of mechanisms, and comparisons with variants or other kinases. Besides these practical motivations, sampling the Abl1 inactivation pathway would additionally be extremely valuable for basic biophysics, as it would further our understanding of enzymatic self-inhibition through intrinsic dynamics.

In addition to exploring the dynamics of the wild-type enzyme, simulations could also be useful for solving longstanding controversies regarding how compound mutants grant Abl1 resistance to inhibitors \cite{Hoemberger2020, Yamamoto2004, Gibbons2011}. Previous studies have found that variants such as Abl1 Glu255Val + Thr315Ile have significantly different activity and binding profiles than wild-type Abl1 \cite{Zabriskie2014, Hoemberger2020}. Crucially, changes in the intrinsic dynamics of Abl1 in response to compound mutations could explain many of the complex effects observed in inhibitor-resistant Abl1 variants.

Beyond the direct impacts of such a study on our understanding of Abl1 kinase and its pharmacology, insights obtained from exploring the molecular mechanisms and propensity for undergoing spontaneous inactivation are potentially transferable to other kinases carrying the DFG motif. Since the DFG motif is nearly universally conserved within the kinome \cite{Modi2021}, this represents significant opportunities for drug discovery and basic biomedicine. Further, other enzymes such as phosphatases carry conserved structural elements such as the WPD motif that have been previously observed to regulate function through backbone torsion mechanisms \cite{Yeh2023} that are analogous to the DFG flip. Insights obtained from our simulations could also be transferable to these systems.  

With this motivation, we have set out to systematically simulate and analyze the dynamics of Abl1 inactivation, specifically those involving the inactivation pathway represented by the DFG flip. To achieve this, we have used the Weighted Ensemble (WE) enhanced sampling approach \cite{Russo2022, Zwier2015, Zuckerman2017}. As the name suggests, enhanced sampling methods are strategies designed to significantly increase the timescales accessible in atomistic simulations, allowing for the effective sampling of physiologically-relevant rare events such as the binding of ligands and large conformational rearrangements \cite{Henin2022}.

In this work, we report thoroughly sampled DFG flip inactivation pathways for a wild-type tyrosine kinase using simulation methods without biasing forces. Through an aggregate simulation time of over 20 $\mu$s, we have collected dozens of uncorrelated and unbiased inactivation events, starting from the lowest energy conformation of the Abl1 kinase core (PDB 6XR6) \cite{Xie2020}. Our results provided us with sufficient data to put forth a broad mechanism for the DFG flip that synthesizes observations made by many previous works regarding kinase structural elements that relate to the DFG flip. 

While previous studies have attributed the inhibitor-resistant phenotype of Abl1 Glu255Val/Lys Thr315Ile to steric clashes with would-be inhibitors \cite{Gibbons2011, Yamamoto2004}, our results reveal that the mechanism for resistance is considerably more complex. Specifically, we have found that the aforementioned variants dramatically shift the conformational preference of apo Abl1, reducing its propensity for undergoing spontaneous inactivation through the DFG flip pathway. This phenotype translates into a major reduction of the population of the DFG-out (inactive) state, the kinase conformation to which Type-II inhibitors preferentially bind. Thus, it follows that drug resistance in the variants we studied stems not just from steric clashes with would-be inhibitors, but also from reduced inhibitor affinity during the conformational search step of binding.

Our results moreover inform predictions about the propensity for the DFG flip in other kinases. By identifying the main electrostatic interactions driving spontaneous inactivation in Abl1, we were able to predict how these interactions might change in Abl1's orthologs including Src, which is known to undergo spontaneous inactivation at much lower rates than Abl1 \cite{Wilson2015}. Crucially, the changes in the N-Lobe negative charge density from Src to Abl1 strongly hint at the atomistic origins of these kinases' contrasting conformational landscapes, as we found that repulsive interactions driven by negatively-charged N-Lobe residues are important for driving Abl1's spontaneous inactivation.


Ultimately, we anticipate that our work will be used to orient the design of the next generation of kinase inhibitors, beyond its importance for basic biophysics.

\section{Results and Discussion \label{results}}

\subsection{Weighted Ensemble Simulations of the DFG Flip\label{westpa}}

Considering the steep free energy barrier of roughly 32 kcal/mol associated with the DFG-in to DFG-out transition in the wild-type Abl1 kinase \cite{Xie2020}, and previous uses of the Weighted Ensemble simulation method for sampling transitions with comparable energy barriers, \cite{Saglam2019} we elected to study the DFG flip in the wild-type Abl1 kinase core (PDB 6XR6, \cite{Xie2020}) using the WE methodology. Applying WE provided us with an abundance of data that enabled us to derive a coherent mechanism for different pathways spanning the transition, and their accompanying probabilities. As a supervised enhanced sampling method, WE employs progress coordinates (PCs) to track the time-dependent evolution of a system from one or more basis states towards a target state. Although weighted ensemble simulations are unbiased in the sense that no biasing forces are added over the course of the simulations, the selection of progress coordinates and the bin definitions can potentially bias the results towards specific pathways \cite{Zuckerman2017}.  Additionally, traditional WEMD simulations do not explicitly enhance sampling along orthogonal degrees of freedom (those not captured by the progress coordinates). In practice, this means that insufficient PC definitions can lead to poor sampling.

In our selection of progress coordinates, we were mindful of this caveat and chose progress coordinates that enhance the sampling while allowing for a variety of different pathways. First, we defined PC1 as the distance between the polar hydrogen of protonated Asp381 and the backbone oxygen of Val299. This choice of progress coordinate is rooted in the observation that the formation of a strong hydrogen interaction between this pair of atoms is likely an important step in the DFG flip in Abl1 kinase \cite{Shan2009}. For PC2, we chose the distance between the gamma carbon of Phe382 and the gamma carbon of Tyr253, which is substantially reduced in DFG-out conformations because the Phe382 side chain approaches the Tyr253 side chain after the flip. PC2 tracks the Phe382 flip, and is pathway-agnostic for that particular event, as the distance between the atom pairs should decrease independently of Phe382 flipping clockwise (passing Thr315) or counterclockwise (passing His361). Finally, we chose the angle formed by Phe382's gamma carbon, Asp381's protonated side chain oxygen (OD2), and Lys378's backbone oxygen as PC3 based on observations from a study that used a similar PC to sample the DFG flip in Aurora Kinase B using metadynamics \cite{Lakkaniga2019}. This angular PC3 should increase or decrease (based on the pathway) during the DFG flip, with peak differences at intermediate DFG configurations, and then revert to its initial state when the flip concludes. Since both an increase or decrease of PC3 accurately tracks flipping events, this progress coordinate is convenient for directing the simulations towards our DFG-out target without imposing significant pathway biases. Our progress coordinates are further illustrated in Figure 2.

For downstream analysis, we used two pseudodihedrals previously defined in the existing Abl1 DFG flip simulation literature \cite{Meng2015} to identify and discriminate between DFG states. The first (dihedral 1) tracks the flip state of Asp381, and is formed by the beta carbon of Ala380, the alpha carbon of Ala380, the alpha carbon of Asp381, and the gamma carbon of Asp381. The second (dihedral 2) tracks the flip state of Phe382, and is formed by the beta carbon of Ala380, the alpha carbon of Ala380, the alpha carbon of Phe381, and the gamma carbon of Phe381. These pseudodihedrals, when plotted in relation to each other, clearly distinguish between the initial DFG-in state, the target DFG-out state, and potential intermediate states in which either Asp381 or Phe381 has flipped.

\begin{figure}[H]
\centering
    \includegraphics[width=500pt]{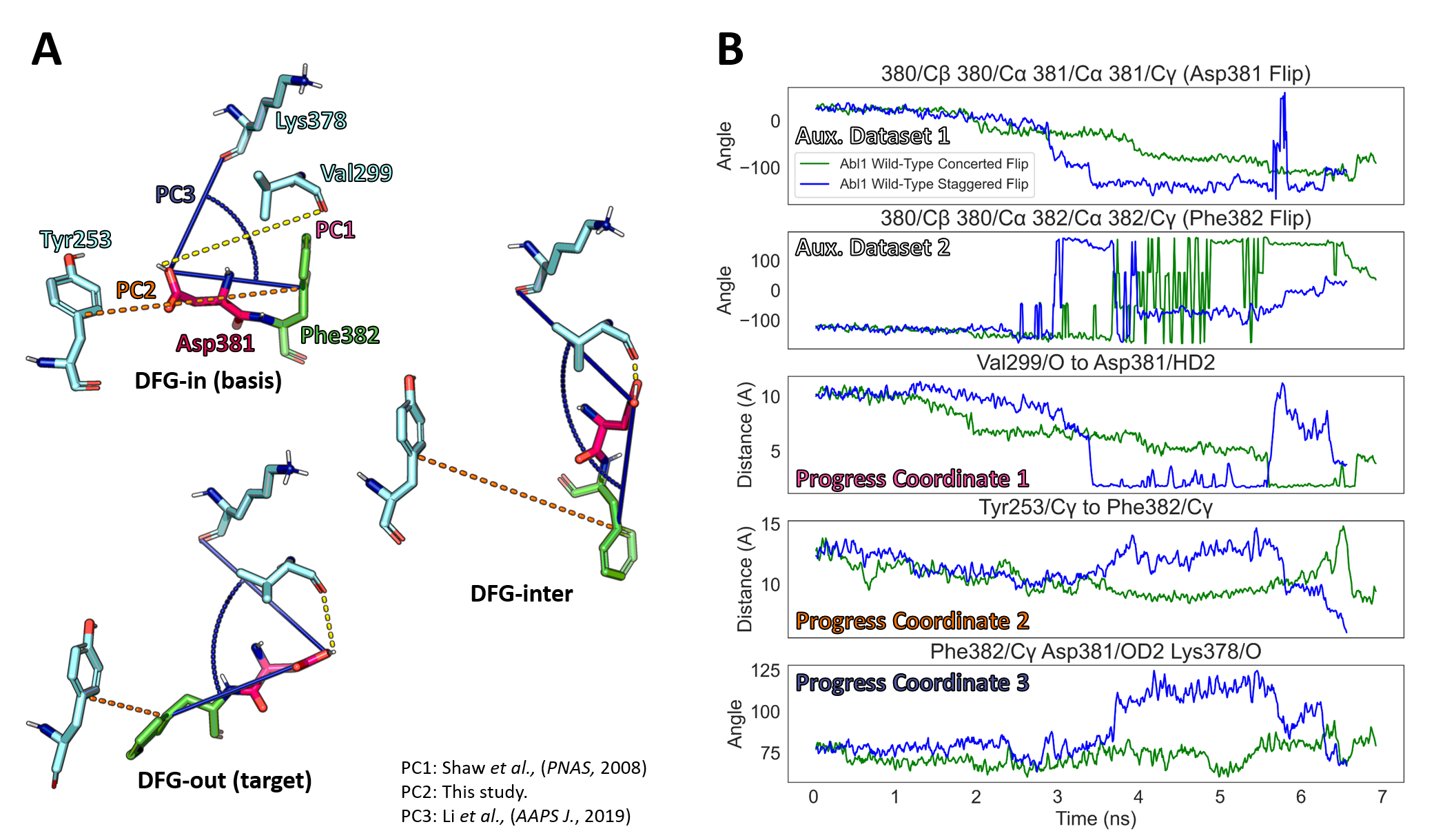}
    \caption{Progress coordinates and auxiliary datasets tracked in the WE simulations of the Abl1 kinase core DFG flip. (A): Structural representation of the observables tracked by each progress coordinate in three contrasting DFG conformational states (DFG-in, DFG-inter, and DFG-out). References that introduced each chosen PC are listed in the bottom right section of the panel. (B): Evolution of auxiliary datasets 1 and 2 (used to track the torsions associated with Asp381/Phe382 flipping) and progress coordinates 1-3 (used to discretize the conformational space populated by the WE simulations and direct walkers towards the DFG-out state). Data were taken from two representative trajectories of the DFG flip in the wild-type Abl1 kinase core simulated with the WE methodology.}
     \label{fig_2}
\end{figure}

Having chosen our set of PCs, we ran WE simulations of wild-type Abl1 starting in the DFG-in conformation with the DFG-out conformation as a target state. Using the apo Abl1 DFG-in and DFG-out structures as references (6XR6 and 6XR7, respectively), we defined bins corresponding to the initial and target states, then used the minimal, adaptive binning (MAB) scheme to further discretize the PC space between the initial and target bins \cite{Torrillo2021}. The MAB scheme expedites sampling by automatically distributing bins on the fly based on the evolution of the system, keeping the usage of computational resources close to constant and maximizing the sampling at bottleneck and boundary PC regions. We ran our WE simulations for 200 iterations, with 8 walkers per bin, 297 bins, totaling a maximum of 2376 segments per iteration, and an iteration time (tau) of 50 picoseconds. Each WE run was repeated two times, totalling three replicates.

Using these parameters, we observed the first DFG flip event in wild-type Abl1 in the first replicate after roughly 157 iterations. Analysis of the concatenated trajectory for this crossing event revealed the existence of a putative intermediate state associated with DFG torsions. For clarity, we will refer to this putative state as DFG-inter. Walkers continued to cross to the DFG-out state until the end of the simulations. To deconvolute the abundance of high-dimensional data obtained from our WE simulations, we ran a variety of analysis methods to measure the relative probability accumulated for walkers in the initial (DFG-in), putative intermediate (DFG-inter), and target (DFG-out) states. The results for this analysis are summarized in Figure 3.

\begin{figure}[H]
\centering
    \includegraphics[width=500pt]{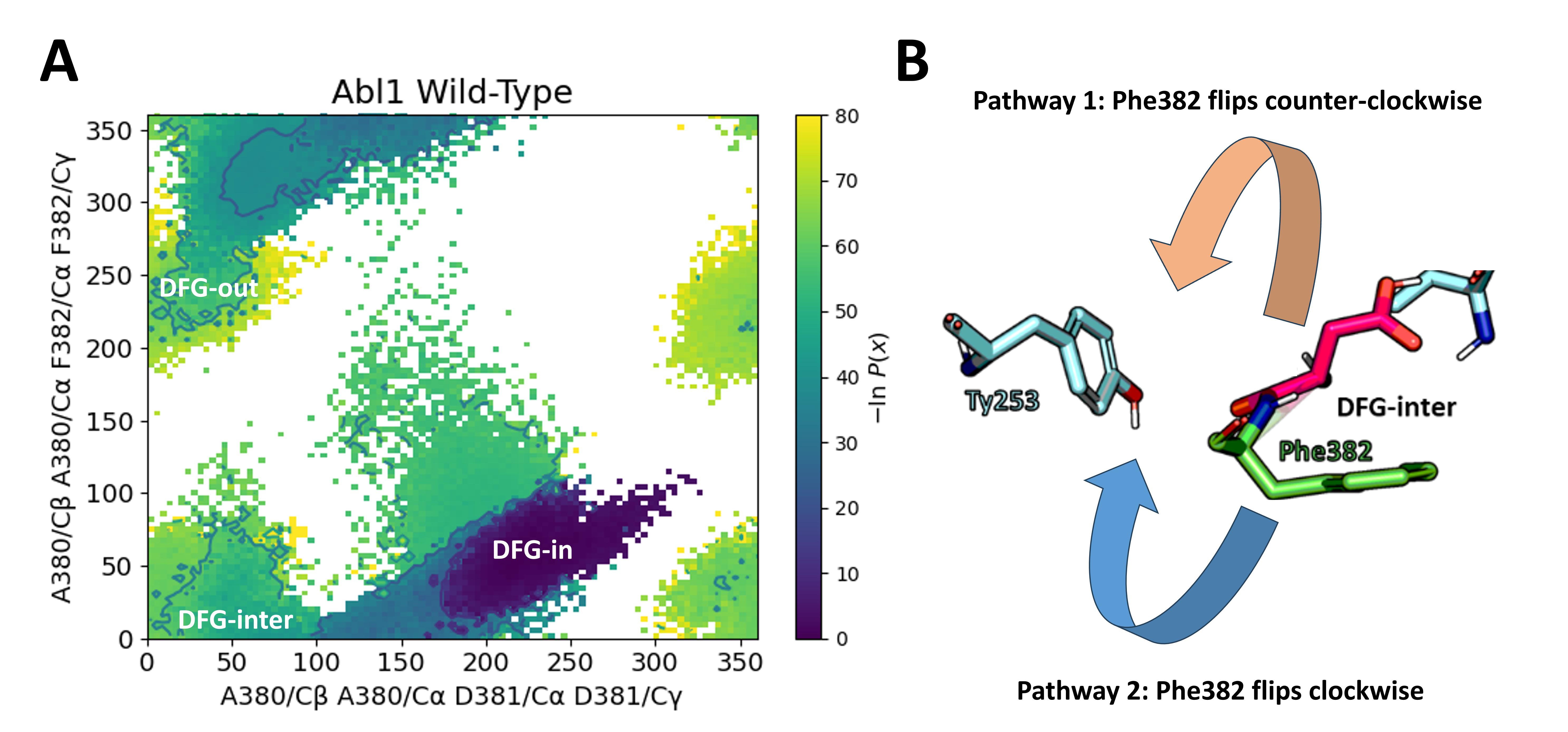}
    \caption{Summary of the conformational space explored in the wild-type Abl1 kinase core WE simulations and pathways identified for the DFG flip. (A): Bi-dimensional projection of the Asp381 and Phe382 torsion angles (tracked by the dihedrals defined in Figure 2) for the entire dataset resulting from the WE simulations. Each dot represents a walker (individual simulation). (B): Visual guide to the key difference between the pathways identified in the WE simulations of the wild-type Abl1 kinase core DFG flip, starting from the common "DFG-inter" DFG intermediate conformation.}
     \label{fig_3}
\end{figure}

Our analysis revealed two predominant pathways for the DFG flip that branch-off at the DFG-inter to DFG-out transition. Specifically, while Asp381 always flips clockwise towards Val299 in all sampled transitions, the flipping of Phe382 is more plastic as it can either flip counter-clockwise after Asp381 nearly completes its flip (staggered DFG flip, Pathway 1) or flip clockwise together with Asp381 (concerted DFG flip, Pathway 2). From our analysis, Pathway 2 has a significantly higher probability than Pathway 1, which is intuitive given that it represents a shorter path to the DFG-out state in terms of the required backbone torsions.

As expected given the significant energy barrier for the flips, the relative probability of walkers reaching the DFG-out state is extremely low (roughly 1.9$\times$ $10^{-23}$). These relative probabilities qualitatively agree with the large expected free energy barrier for the DFG-in to DFG-out transition (\~32 kcal/mol).

Despite our efforts to maximize pathway resolution, it is impossible to completely remove the biasing effects of progress coordinate choice. As there are an astronomical number of possible pathways along which a macromolecule can undergo torsion, we could still miss many with our PC choice. Given that limitation, we constrain the scope of the following discussion to the predominant pathways discussed above. Further studies using specialized force fields and alternative enhanced sampling schemes would be important to address the potential limitations of our approach.

\subsection{Molecular Mechanism of the DFG Flip\label{mechanism_wt}}

Making use of the wealth of structural information generated from our weighted ensemble simulations of the DFG flip in Abl1, we sought to determine a general molecular mechanism involving N-Lobe structural elements for the DFG flip pathways sampled in our simulations. To do so, we selected sample trajectories containing the successful flip event (DFG-in to DFG-out) along the most frequent pathway (concerted flip), and analyzed the evolution of key structural elements thought or known to be important for the DFG flip. The results of this analysis are reported in Figure 4, and the selected trajectories accompanied by their topology files are available in the project's GitHub repository \cite{abl1_dfgflip_westpa_rawdata}.

\begin{figure}[H]
\centering
    \includegraphics[width=500pt]{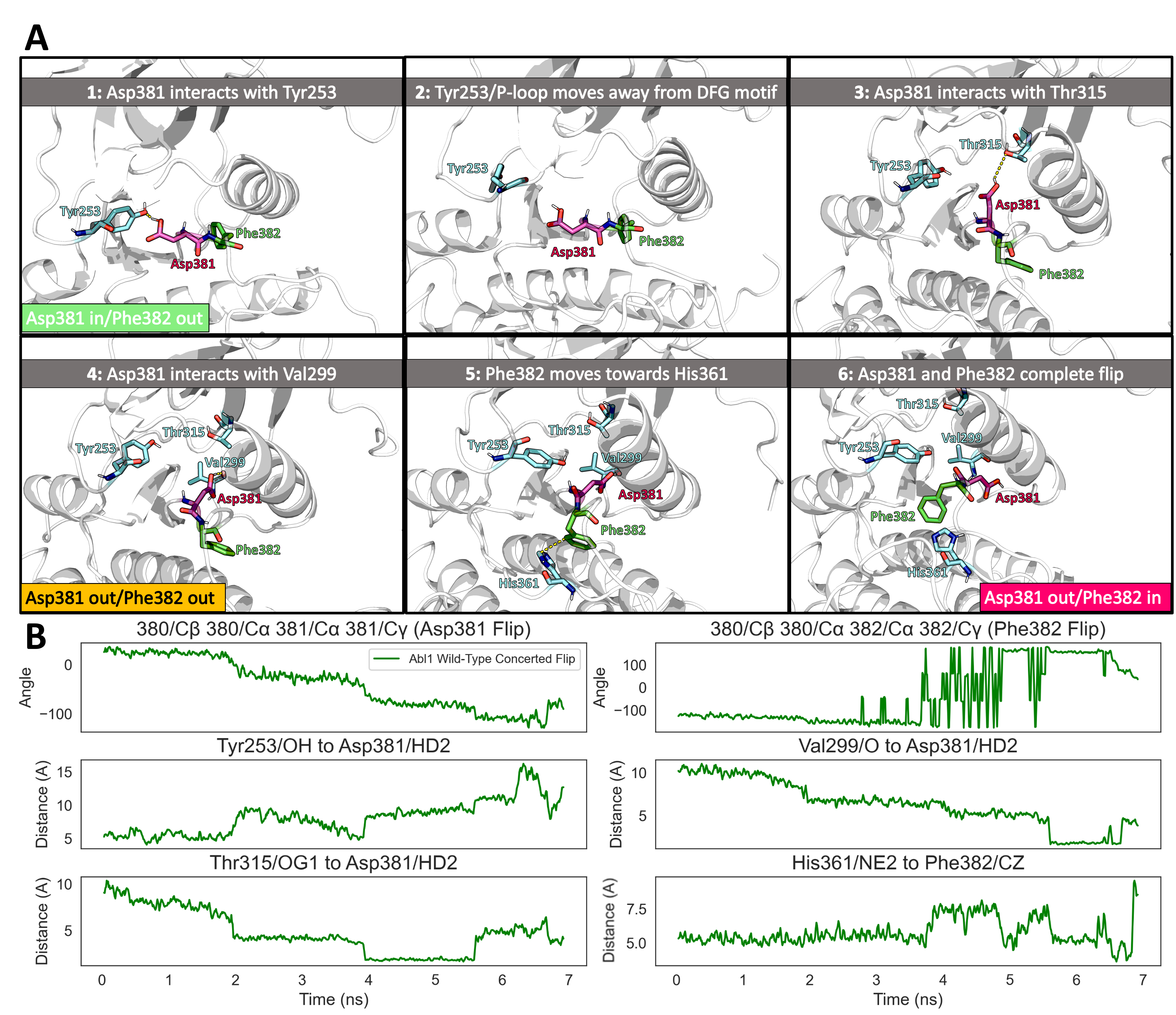}
    \caption{Detailed analysis of a successful DFG flip event in wild-type Abl1 kinase simulated with the weighted ensemble method. (A) Representative frames from the selected concerted DFG flip trajectory arranged in a timeline showcasing important interactions and rearrangements for the flip. Frames are ordered from left to right, top to bottom. (B) Evolution of selected observables shown in (A) during the DFG flip trajectory.}
     \label{fig_4}
\end{figure}

In the concerted pathway, the first stage of the flip corresponds to the partial flips of Asp381 and Phe382 (DFG-in to DFG-inter), and the second corresponds to the completion of the Asp381 flip in concert with the completion of the Phe382 flip (DFG-inter to DFG-out). In the former, the side chain of Asp381 makes its way into the vicinity of the initial position of Phe382's side chain through a relatively slow process (spanning 5 ns in the selected trajectory). First, P-Loop fluctuations break the interaction between the Asp381 polar hydrogen and the P-Loop Tyr253's oxygen, driving Asp381 towards the solvent-accessible gap directly below Thr315 \cite{10.145/118455}. Upon entering that gap, the polar hydrogen of protonated Asp381 forms a brief hydrogen interaction with Thr315's side chain, which is a weak hydrogen acceptor in this context. The formation of this interaction immediately provides insights into the effects of Thr315 mutations on the propensity for the DFG flip. By analyzing the distributions of the distances between the atoms involved in the Thr315-Asp381 interaction in our WE simulations during DFG torsions, we observed a strong correlation between the strength of the Thr315-Asp381 interaction and DFG conformations (Supplementary Figure 1).

Next, the side chain of Asp381 crosses over towards Phe382's initial position, forming a stable hydrogen bond with the backbone oxygen of Val299. This interaction is known to be important for the flip, and was previously described as a mechanism for the significantly increased propensity for the DFG flip in protonated Asp381 \cite{Shan2009}. The formation of this interaction marks the end of the first step of the proposed mechanism for the concerted DFG flip pathway, as now the side chains of Asp381 and Phe382 shifted from their original positions and both point towards the \(\alpha\)C-Helix side of the N-Lobe (DFG-inter). Importantly, since the side chain of Asp381 in this DFG-inter conformation points away from the ATP-binding site, it is unlikely that this conformation is capable of productive phosphorylation.

Eventually, changes in the \(\alpha\)C-Helix orientation and distance to the DFG motif allow Asp381 to complete the flip, breaking the interaction with Val299. In a concerted fashion, Phe382 also moves down towards the HRD motif, a highly-conserved segment in protein tyrosine kinases which is involved in catalysis. This movement is facilitated by the absence of steric clashes with the \(\alpha\)C-Helix due to its previous reorientation. Concurrently, Phe382's side chain moves towards the phosphorylation site in a clockwise motion, eventually forming favorable hydrophobic contacts and $\pi$-stacking interactions with Tyr253, completing the flip. 

Previous studies simulating the DFG flip have found that brief breaks in the salt bridge formed by Lys271 and Glu286 coincide with Asp381 and Phe382 flipping \cite{Narayan2020}. Overwhelmingly conserved among kinase domains \cite{Modi2021}, the Lys271-Glu286 interaction plays a significant role in stabilizing the N-Lobe architecture by anchoring the \(\alpha\)C-Helix. The correlation between Lys271-Glu286 salt bridge occupancy and DFG torsions was also captured in concerted DFG flips from our WE simulations, as the distances between the atoms involved in forming the salt bridge are significantly larger in intermediate DFG conformations than they are in the DFG-in or DFG-out states (Figure 5). These results provide further evidence for the role of the Lys271-Glu286 salt bridge in the DFG flip mechanism.

\begin{figure}[H]
\centering\includegraphics[width=500pt]{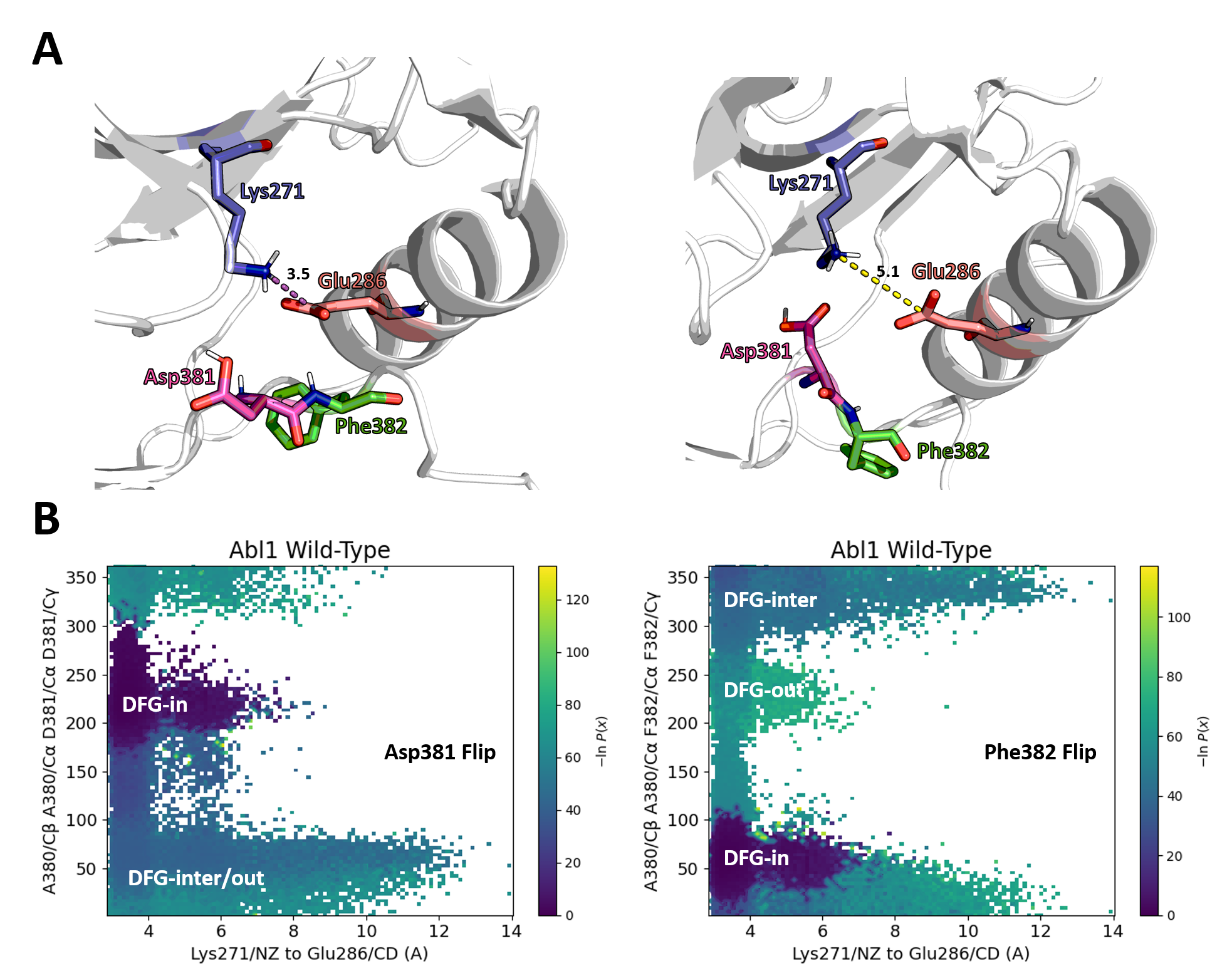}
    \caption{Correlation between the Lys271/Glu286 salt bridge and DFG torsions in WE simulations of the wild-type Abl1 kinase core DFG flip. (A): Visual representation of the distances between the terminal carbon of Glu286 and the charged nitrogen of Lys271 in two distinct DFG conformations (starting conformation on the left; and a DFG flip intermediate conformation on the right). (B): Bi-dimensional projection of the correlation between Lys271/Glu286 distances and Asp381 (left) or Phe382 (right) torsions for the entire WE simulation dataset. Each dot corresponds to an individual simulation (walker), and dots are colored based on the inverse of their relative probabilities. Starting, intermediate, and target states are labeled in the plots.}
     \label{fig_5}
\end{figure}

The comparatively rarer ``Staggered DFG Flip'' pathway follows a mechanism that conserves many of the features of the previously described ``Concerted DFG Flip''pathway (Supplementary Figure 2). The crucial difference between the mechanisms is that, in the staggered flip, Phe382's torsions are stabilized in a near-native conformation while Asp381 maintains its hydrogen bond with Val299 (from 3.5 ns to 6.5 ns in the sample trajectory). This intermediate conformation poises Phe382 to flip counter-clockwise, crossing the same path as Asp381, but in reverse. Interestingly, the Lys271-Glu286 salt bridge remains stable during staggered DFG flips (Figure 5B), which further substantiates the reduced probability for the flip to happen through this mechanism.

In addition to the data presented above, we also measured the distribution and probabilities of observables important for the DFG flip during entire WE simulations (containing thousands of individual walkers). These data for PC1 through PC3 are analyzed in Supplementary Figure 3.

Our model accounts for the role of multiple structural elements in Abl1 that were previously suggested to be important for the DFG flip, such as the Val299-Asp381 hydrogen interaction and the Lys271-Glu286 salt bridge, while also newly highlighting the importance of contacts with Tyr253 and of the hydrogen interaction with Thr315 (which, as mentioned, provides interesting insights into the effects of the Thr315Ile mutation). During all simulated flip events, Asp381 flips clockwise and its protonated side chain crosses through a series of interaction events before completing the flip, often taking many nanoseconds to go from one interaction event to the other. The dynamics of Phe382, on the other hand, are substantially faster once they begin, because Phe382 makes no detectable stable interactions with other residues during the transition.

As a crucial caveat, a clockwise flip of Phe382 (in the concerted flip pathway) puts its side chain in the vicinity of that of His361, which could potentially stabilize it in an intermediate conformation through $\pi$-$\pi$ stacking interactions. Indeed, the angles and distances of the Phe382-His361 side chain observed in trajectories where Phe382 flips clockwise could be indicative of $\pi$-$\pi$ stacking. The approximations inherent to the force field used in this simulation (CHARMM36M, \cite{Huang2016}) prevent us from simulating these interactions directly, and could lead to this pathway being underrepresented in the resulting ensemble.

\subsection{WESTPA Simulations of the DFG Flip in Abl1 Variants\label{westpa_muts}}

As previously discussed, a collection of mutations in the Abl1 kinase core are known drivers of inhibitor resistance \cite{Hoemberger2020, Yamamoto2004, Azam2008, Lyczek2021}. Among these variants, mutations to Thr315 and Glu255 are of particular concern because of their high prevalence in patient tumor samples and their capacity for inducing drug resistance even against last-generation inhibitors such as Ponatinib when manifested together \cite{Zabriskie2014} (Supplementary Table 1). 

Thr315Ile, known as the gatekeeper mutation due to its side chain steric clashes with Imatinib that hinder binding, is of particular concern for Abl1 inhibition. In contrast with those of Thr315Ile, the structural consequences of Glu255 mutations to valine or lysine are significantly less understood. Since the side chain of Glu255 usually points away from the ATP-binding site, steric clashes with inhibitors are not as likely, and allosteric effects such as decreases in P-Loop stability or Lys271/Glu286 salt bridge formation are suspected to cause inhibition \cite{Zabriskie2014}. While Ponatinib and other last-generation inhibitors can successfully inhibit Abl1 Thr315Ile, the combination of this variant with Glu255 mutations sharply decreases Ponatinib's affinity \cite{Zabriskie2014}, representing a continuing public health risk.

To interrogate the molecular mechanisms of drug resistance in Abl1's compound variants Glu255Val+Thr315Ile and Glu255Lys+Thr315Ile, we repeated our WE workflow on these variants of Abl1 in the DFG-in state as the starting configuration (see the Methods Section for details on how the mutations were modeled). The results of these simulations are summarized in Figure 6.

\begin{figure}[H]
\centering
\includegraphics[width=450pt]{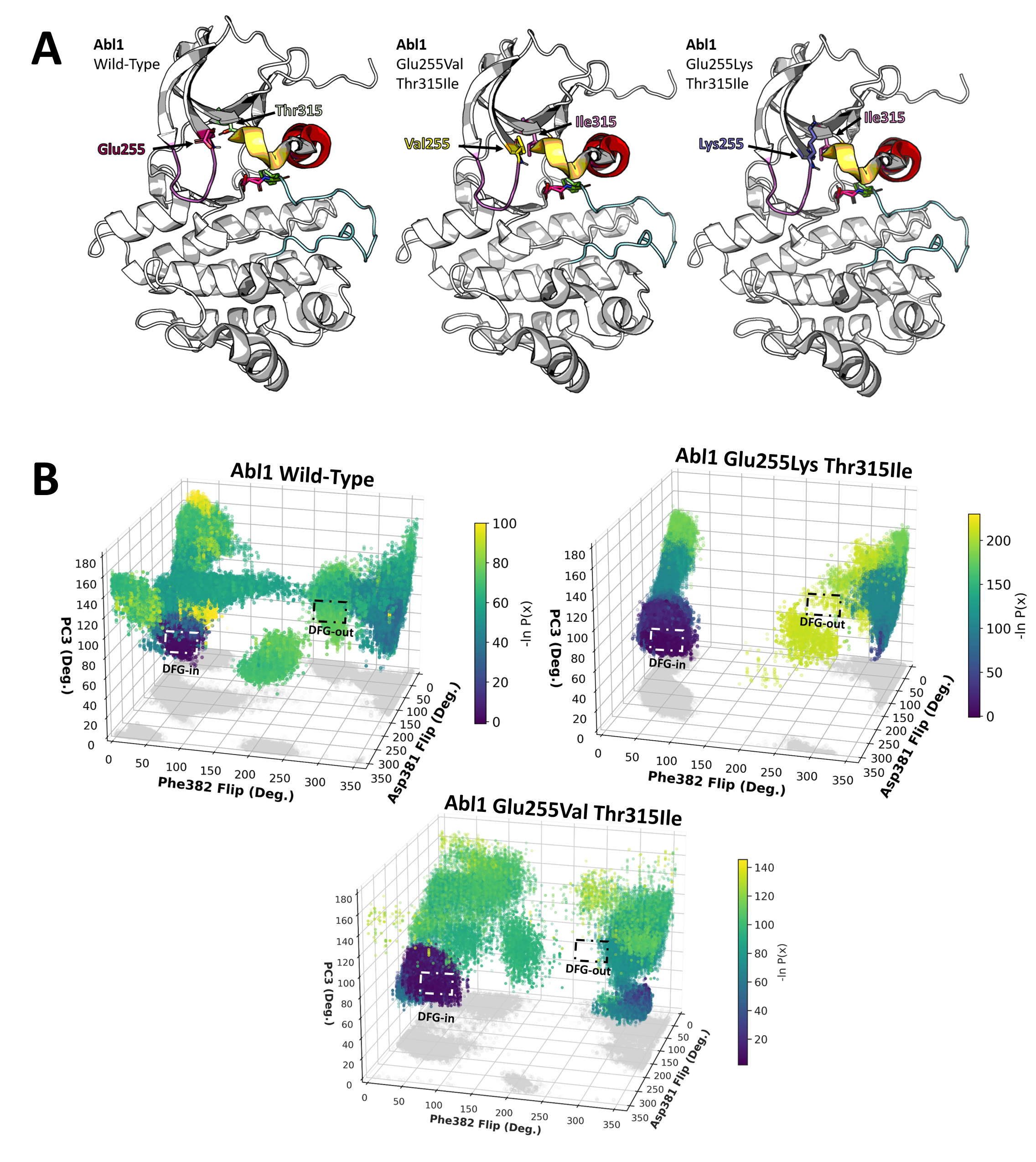}
    \caption{Results of weighted ensemble simulations of the DFG flip in the Abl1 kinase core and Abl1's drug-resistant variants. (A) Structural representation of the Abl1 kinase core and the positions of wild-type residues Glu255 and Thr315 and their variants, Val/Lys255 and Ile315. For ease of visualization, key residues are represented as sticks (Asp381, red; Phe382, green; Glu255, orange; Thr315, cyan; Val255, yellow; Lys255, blue; Ile315, magenta). Additional elements important for kinase function and involved in the DFG flip are colored in pink (P-Loop, residues 249-255), orange (B-Loop, residues 275-278), red (\(\alpha\)C-Helix, residues 280-293), and cyan (activation loop, residues 385-400). (B) Tri-dimensional projection of the values of the torsion angles used to define and visualize the state of the DFG motif in Abl1 kinase in the weighted ensemble simulations (X and Y) and of Progress Coordinate 3 (PC3, Z). Color bars define the probability associated with each walker, and are a surrogate for the free energy barrier between states.}
     \label{fig_6}
\end{figure}

DFG flip events in simulations of the Abl1 compound mutants occurred at substantially lower probabilities than in the wild-type enzyme (Supplementary Table 2). Further, simulations of wild-type Abl1 sampled significantly broader regions of conformational space than simulations of the mutants. 

Importantly, these simulations were run with the exact same parameters and for the exact same length (200 iterations) as the wild-type simulations. We also report the evolution, distribution, and probabilities of observables used as progress coordinates or to define the DFG flip in the mutant WE simulations, which can be found in Supplementary Figures 4 and 5.

In the Glu255Lys + Thr315Ile simulations, the Abl1 kinase core proceeds relatively quickly (in fewer than 100 WE iterations) to the DFG-inter state, in which Asp381 has crossed over to make the interaction with Val299, but the DFG flip is ``halted'' as the kinase remains remarkably stable in this conformation before eventually completing the flip through the Asp381 clockwise/Phe382 clockwise pathway. Compared to the wild-type simulations, considerably fewer walkers complete the flip, and those that do so, do with extremely low probabilities.

Crucially, the fact that the probabilities for both double mutants to complete the DFG flip are significantly less than those of the wild-type enzyme suggests that the free energy barriers for the flip are sharply increased in the compound mutants. This increase could be caused by a variety of mechanisms, including stabilization of the DFG-inter conformation, but its macroscopic outcome is a significantly reduced preference for binding Type-II inhibitors such as Imatinib or Ponatinib (which bind preferentially to the DFG-out conformation).

As an important caveat, it is unlikely that the DFG flip free energy barriers of over 70 kcal/mol estimated for the Abl1 drug-resistant variants quantitatively match the expected free energy barrier for their inactivation. Rather, our approximate free energy barriers are a symptom of the markedly increased simulation time required to sample the DFG flip in the variants relative to the wild-type, which is a strong indicator of the drastically reduced propensity of the variants to complete the DFG flip. Although longer WE simulations could allow us to access the timescales necessary for more accurately sampling the free energy barriers associated with the DFG flip in Abl1's drug-resistant compound mutants, the computational expense of running WE for 200 iterations is already large (three weeks with 8 NVIDIA RTX3900 GPUs for one replicate); this poses a logistical barrier to attempting to sample sufficient events to be able to fully characterize how the reaction path and free energy barrier change for the flip associated with the mutations. Regardless, the results of our WE simulations resoundingly show that the Glu255Lys/Val and Thr315Ile compound mutations drastically reduce the probability for DFG flip events in Abl1.

From analyzing the ensembles of trajectories of wild-type vs. drug-resistant Abl1 variants resulting from the WE simulations, we observe that walkers in the Glu255Lys Thr315Ile variant encounter a significant bottleneck surmounting the barrier between the DFG-in and DFG-inter states. We therefore selected a representative trajectory for this ''Frustrated DFG Flip'' event for the downstream analyses described in the following sections. Similarly, walkers in the Glu255Val Thr315Ile simulations halt at the DFG-inter state more frequently than in the wild-type. Thus, we selected a representative trajectory for this ''Half DFG Flip'' event as well.

\subsection{Electrostatic Interactions Involved in the Flip\label{electrostatics}}

Considering our previous findings regarding the wild-type Abl1 kinase core DFG flip and the impact of the Glu255 mutations on this flip, we sought to more thoroughly analyze the role of Glu255 in successful (concerted or staggered), half or frustrated DFG flip simulations of wild-type Abl1 and its drug-resistant variants. To accomplish this, we used the representative trajectories selected from the weighted ensemble simulation results described above, and analyzed the formation or breaking of electrostatic interactions involving Glu255 in these trajectories. As previously mentioned, these sample trajectories are deposited in this study's GitHub repository \cite{abl1_dfgflip_westpa_rawdata}. The results for a subset of potentially important interactions are summarized in Figure 7.

\begin{figure}[H]
\centering
\includegraphics[width=450pt]{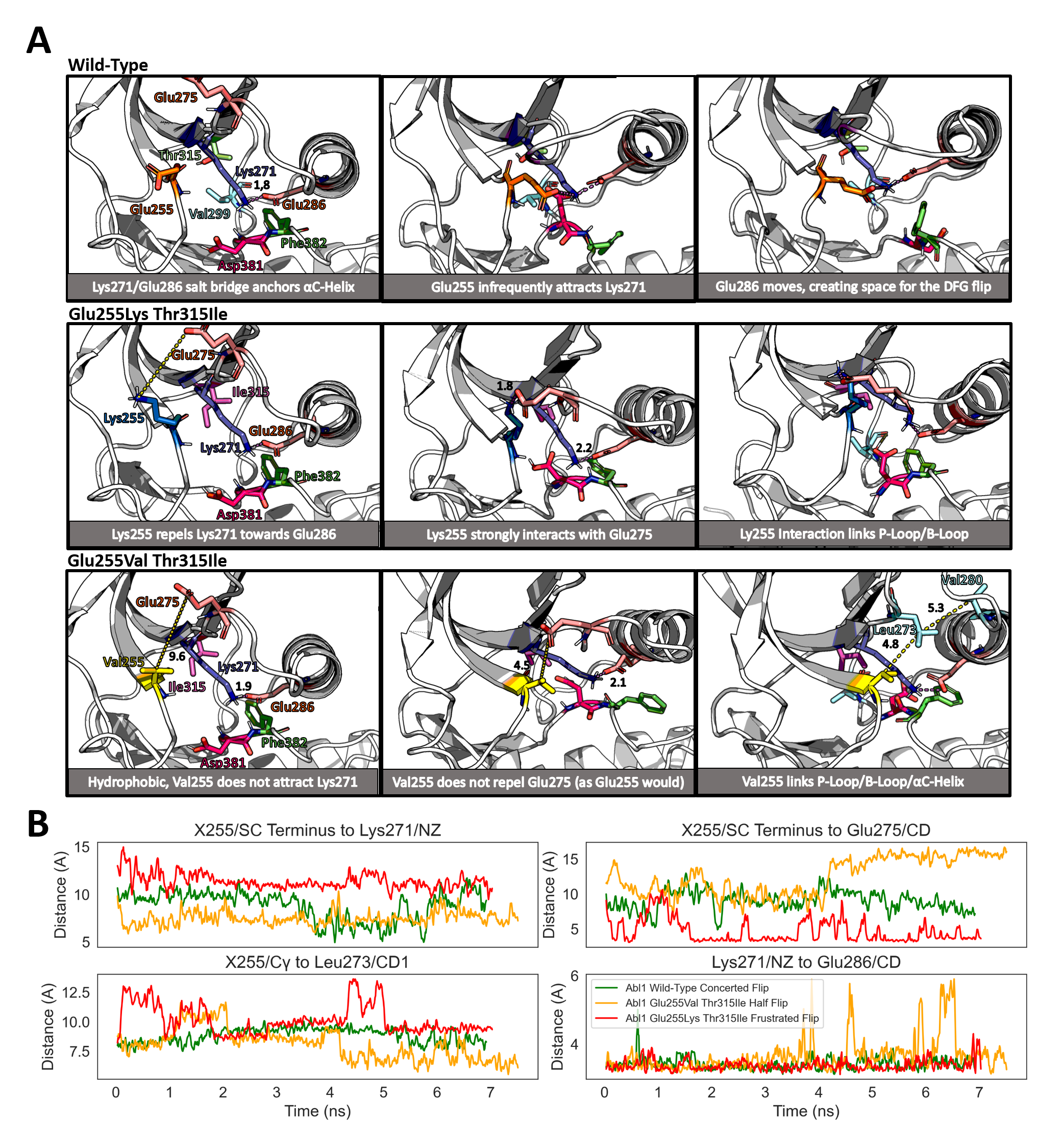}\caption{Contributions of different electrostatic interactions to DFG flip events in Abl1 kinase and its drug-resistant variants. (A) Representative frames showcasing relevant interactions and residue positions for complete (top), half (middle), and frustrated (bottom) DFG flip events taken from the weighted ensemble simulations of each Abl1 variant. Distances between reference atoms are depicted as yellow dotted lines, which are changed to magenta dotted lines if residues form a salt bridge interaction. Distances are given in angstroms. (B) Evolution of selected observables showcased in (A) from representative trajectories for each variant.}
     \label{fig_7}
\end{figure}

Although we limit the following discussion to selected trajectories representative of each ensemble for the sake of clarity, we have also analyzed the distributions and probabilities of DFG-flip-related observables from the entire WE dataset for each simulation. Rationales for these observables are provided in Supplementary Table 3, and their distributions and probabilities are illustrated in Supplementary Figures 6 and 7.

In our simulations, Glu255 infrequently interacts with Lys271 (Supplementary Figures 6 and 7). This interaction could briefly weaken the salt bridge between the side chains of Lys271 and Glu286, increasing the mobility of the \(\alpha\)C-Helix, and potentially creating space for the flip. Although the approximations intrinsic to additive force fields such as the one used in this study (CHARMM36m, \cite{Huang2016}) do not allow for the direct modeling of the effects of polarization and its structural consequences, the displacement of the side chain of Lys271 towards Glu255 due to fixed charge electrostatics is captured in our simulations (Figure 7B), and is a proxy for the polarization effects of the Glu255-Lys271-Glu275 triplet of interactions.

Further, the side chain of Glu255 is also in close proximity to Glu275's side chain, resulting in a repulsion between this pair of negatively-charged moieties. This repulsion could also play a role in causing the \(\alpha\)C-Helix to move and create space for the DFG flip. In contrast, both the Glu255Lys Thr315Ile and Glu255Val Thr315Ile Abl1 variants lack the capacity to form this interaction with Lys271 or repel Glu275. Importantly, the \(\alpha\)C-Helix is more mobile in wild-type simulations than in either compound variant (Supplementary Figures 6 and 7).

Positively-charged, the side chain of Lys255 forms a very stable interaction with Glu275's side chain in the Glu255Lys Thr315Ile variant, linking the end of the P-Loop to the middle of the B-Loop. This interaction shifts the P-Loop in and effectively creates an additional anchor for the \(\alpha\)C-Helix, stabilizing it in conformations even closer to the DFG motif and preventing the dynamics necessary for creating space for the DFG flip (Supplementary Figure 6). Indeed, we observe that the distances between the Lys255-Glu275 residue pair (in the variant) are significantly shorter than the distance between the Glu255-Glu275 pair (in the wild-type) for most of our WE walkers, and that both the \(\alpha\)C-Helix and B-Loop are considerably more stable in the variant (Supplementary Figure 6).

The Glu255Val variant also significantly affects the N-Lobe architecture. Hydrophobic, the side chain of Val255 remains stable for the majority of the representative simulation of this variant and does not disrupt the Lys271-Glu286 salt bridge through direct interactions with either side chain. However, a significant and lasting disruption is observed in the latter half of the trajectory (Figure 7B, from 3.5 to 7.3 ns), which is attributed to a large conformational rearrangement of the Abl1 Glu255Val Thr315Ile N-Lobe, involving the formation of a network of hydrophobic interactions among the side chains of Val255 (mutated from Glu255), Leu273, and Val280. This network creates a stable link between the end of the P-Loop (Val255), the B-Loop (Leu273), and the start of the \(\alpha\)C-Helix (Val280). The consequence of this link is a remarkably stable \(\alpha\)C-Helix orientation after the conformational change, in close proximity to the DFG motif, and a shifted P-Loop that would clash with Ponatinib. In summary, although the Val255 mutation disrupts the Lys271-Glu286 salt bridge, this disruption leads to a conformation in which the space surrounding the DFG motif is even more constricted due to the P-Loop/B-Loop/\(\alpha\)C-Helix rearrangement and lack of repulsion with Glu275. A comparison between the P-Loop and \(\alpha\)C-Helix stability and the distribution and probabilities of either distance pair (Val255-Leu273, for the variant; and Glu255-Leu273, for the wild-type) over the entire WE simulations corroborate the trends we have discerned from this sample trajectory (Supplementary Figure 7).

Together, these results suggest that P-Loop/B-Loop/\(\alpha\)C-Helix interactions mediated by residue 255 assume a significant role in modulating the likelihood of DFG flip events in Abl1 kinase, and the substitution of Glu255 impacts the dynamics of multiple N-Lobe elements whose dynamics are strongly correlated with successful flips of Asp381 and Phe382.

\subsection{Distance Analysis}

To directly interrogate the effect of the N-Lobe's spatial organization on DFG flip events in Abl1, we defined a set of  reference measurements that summarize key properties of the N-Lobe (such as the distance between \(\alpha\)C-Helix residues and the DFG motif) and measured the evolution and distribution of these properties in the representative trajectories described in previous sections, contrasting the wild-type's successful flip in the predominant pathway (concerted flip) with trajectories from either drug-resistant variant. The specific definitions of the distances tracked along with measurements for the Glu255Lys Thr315Ile variant trajectory are provided in Figure 8. Distributions and probabilities for observables measured from the collection of walkers resulting from the WE simulations are described in Supplementary Figures 6 and 7. We also performed this analysis for concerted vs. staggered DFG flips in the wild-type simulations, with the resulting data shown in Supplementary Figure 8.

\begin{figure}[H]
\centering
    \includegraphics[width=450pt]{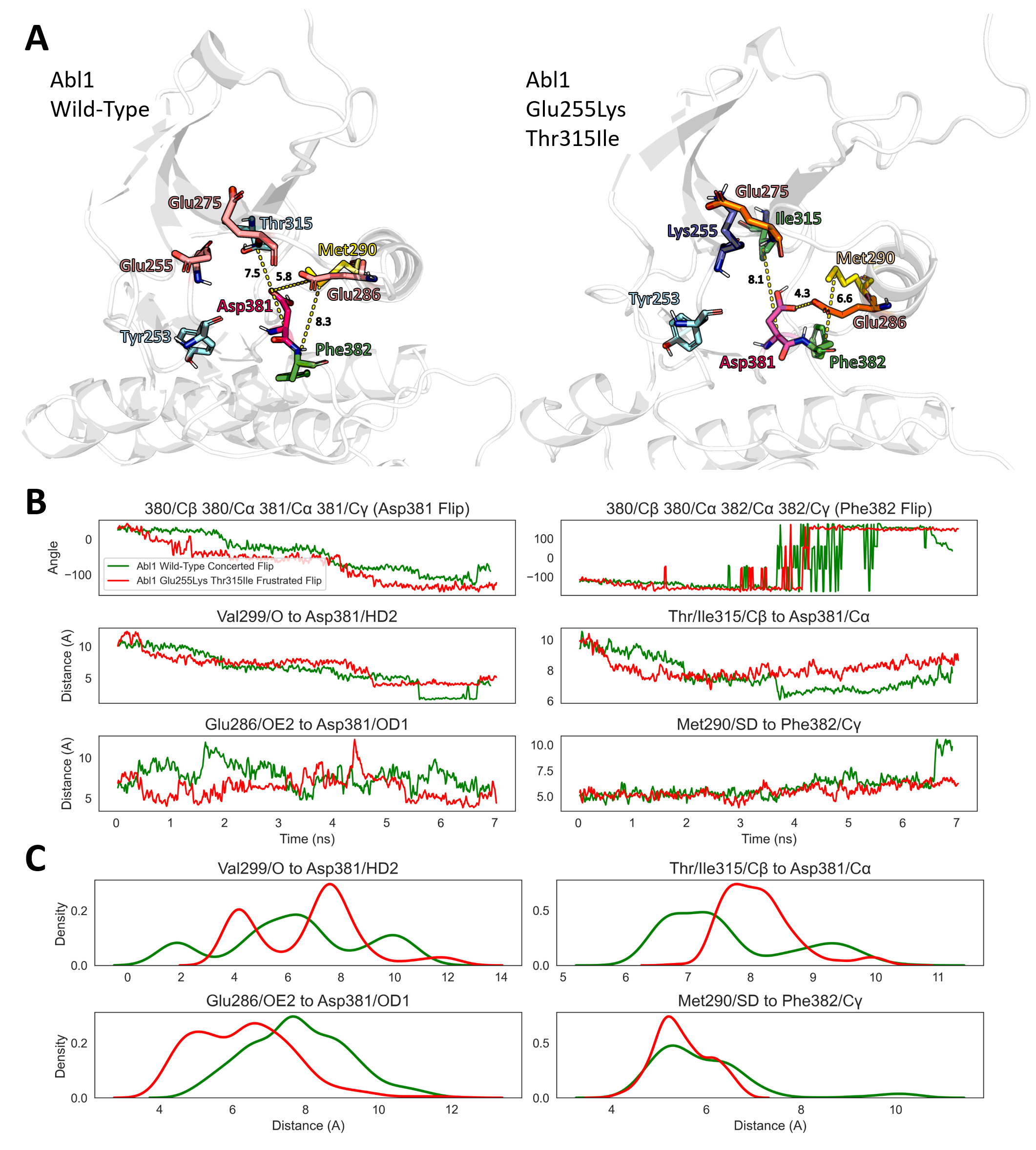}
    \caption{Comparison between intermediate conformations adopted by the Abl1 kinase core and a drug-resistant variant during a DFG flip event. (A) Structural representations of the (left) wild-type and (right) Glu255Lys Thr315Ile Abl1 kinase cores. Residues relevant for inducing or stabilizing either conformation are shown as sticks, accompanied by distance measurements (in Angstroms, shown as yellow dotted lines) to highlight the differences between conformations. (B) Evolution of distances and angles shown in A during a successful DFG flip trajectory (for wild-type Abl1) or a frustrated DFG flip trajectory (for the Glu255Lys Thr315Ile variant). (C) Kernel density estimation of distances shown in B. Trajectories were chosen as representatives from the weighted ensemble simulation results.}
     \label{fig_8}
\end{figure}

In-line with our previous discussion, we observe that, in a successful concerted flip trajectory, the \(\alpha\)C-Helix references (Glu286 and Met290) move away from the DFG motif in both halves of the proposed pathway: immediately before Asp381 begins flipping (1.5 to 3 ns) and again during the Phe382 flip (5 to 7 ns). Glu286 and Met290 were chosen as distance references for the \(\alpha\)C-Helix because their side chains create steric clashes preventing either Asp381 or Phe382 from flipping in the frustrated flip trajectory (Figure 8B). Although the \(\alpha\)C-Helix does not directly clash with Asp381 in its DFG-in conformation, this lateral expansion is important to create space for Asp381's side chain to approach Val299's backbone oxygen and form the hydrogen bond previously observed to be an important step in the flip. 

Additionally, the distance between the beta carbon of Thr315 and Asp381's alpha carbon remains consistently large (>8 \r{A}) until Asp381's polar hydrogen is close enough to interact with Thr315's side-chain oxygen. This separation is important as it prevents unfavorable contacts between the side chains and allows Asp381 to flip mostly unhindered. After the interaction with Thr315 breaks, Asp381 proceeds to form the hydrogen interaction with Val299, after which the flip is complete.

In concert with the previously described Asp381 motions, torsions in Phe382's backbone cause its side chain to move under the \(\alpha\)C-Helix and become mostly solvent-exposed, which creates an unfavorable environment for the hydrophobic side chain. Additional torsions due to Asp381 movements and the unfavorable environment push Phe382's side chain into the vicinity of His361 in the catalytic loop, and eventually to the starting position of Asp381 in the DFG-in state.

Taken together and in the order they are observed, these dynamics represent an intricate dance between the DFG motif and its surroundings as either Asp381 and Phe382 negotiate for space across the crowded environments encountered during a DFG flip event. In stark contrast with the patterns described above, the frustrated DFG flip trajectory taken from the Imatinib-resistant Abl1 Glu255Lys Thr315Ile variant shows a significantly more constricted N-Lobe, with fewer opportunities for a successful DFG flip. In this trajectory, the \(\alpha\)C-Helix to DFG motif distances remain consistently smaller than in the wild-type, as evidenced by the kernel density estimation (Figure 8C), resulting in a significant reduction in the available space for Asp381 or Phe382 to flip. 

Although the side chain of Asp381 does move towards Val299, that movement never consummates into a stable interaction in this trajectory. This deviation from the wild-type pathway is explained by analyzing the distance between the beta carbon of Ile315 and Asp381's alpha carbon, which is significantly smaller than its counterpart in the wild-type  prior to the Asp381-Thr315 interaction, and remains so for the entire trajectory. Bulkier than threonine and incapable of accepting a hydrogen from Asp381, the close proximity of Ile315's side chain forms a notable obstacle for the first half of our proposed mechanism.

In light of the drastic changes in the charged interactions arising from the Glu255Lys mutation described in the previous section, these contrasting distributions provide significant support for the idea that the Glu255Lys Thr315Ile variant begets its drug-resistant phenotype by limiting the probability of successful DFG flips in Abl1, increasing the enzyme's preference for the active conformation, in addition to steric clashes with potential ligands.

The representative trajectory taken from the Glu255Val Thr315Ile Abl1 simulations also shows significant differences with the wild-type trajectory. In this comparison, two properties show signs of being detrimental to a successful DFG flip. First, the \(\alpha\)C-Helix distance to Phe382 is consistently smaller than in the wild-type reference, reducing  the space necessary for Phe382 to flip. Second, the hydrophobic Ile315 remains close to the DFG motif across the entire trajectory, presenting an obstacle for completion of the flip in the same way as discussed for the Glu255Lys Th315Ile variant (Figure 9).

\begin{figure}[H]
\centering
    \includegraphics[width=450pt]{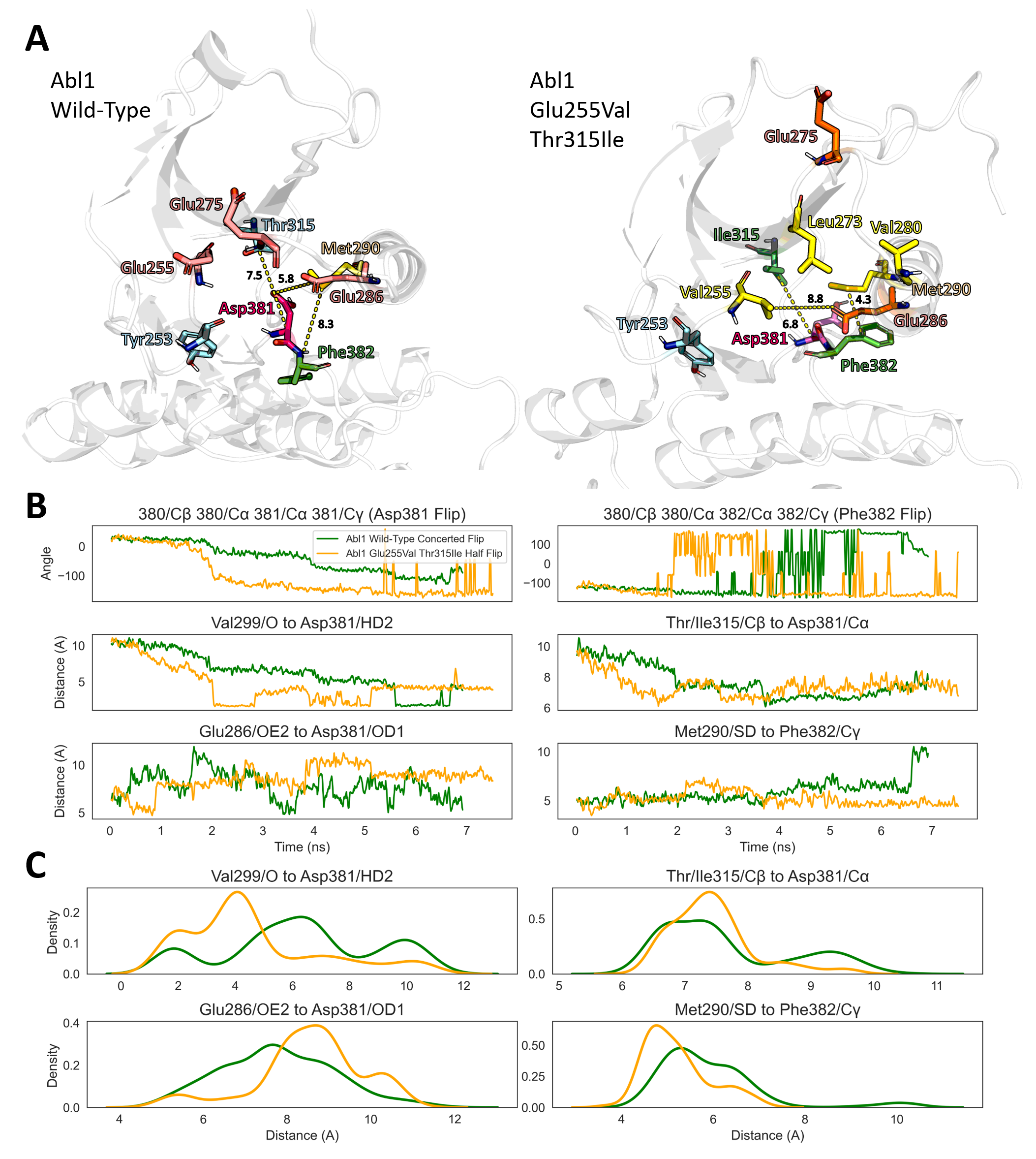}
    \caption{Comparison between intermediate conformations adopted by the Abl1 kinase core and by a drug-resistant variant during a DFG flip event. (A) Structural representation of the (left) wild-type and (right) Glu255Val Thr315Ile Abl1 kinase cores. Residues relevant for inducing or stabilizing either conformation are shown as sticks, accompanied by distance measurements (in Angstroms, shown as yellow dotted lines) to highlight the differences between either conformation. (B) Evolution of distances and angles shown in A during a successful DFG flip trajectory (for wild-type Abl1) and for a half DFG flip trajectory (for the Glu255Val Thr315Ile variant). (C) Kernel density estimation of distances shown in (B). Trajectories were chosen as representatives from the weighted ensemble simulation results.}
     \label{fig_9}
\end{figure}

As discussed in the previous section, the Glu255Val mutation leads to a trajectory in which Abl1's N-Lobe reorients significantly after the DFG-inter state is reached, as hydrophobic interactions are formed between Val255, Leu273, and Val280, which require a considerable shift of the P-Loop, B-Loop, and \(\alpha\)C-Helix. This rearrangement happens at roughly the halfway point of the trajectory (4 ns), and the resulting conformation is stable for the remainder of the the simulation. In this alternative conformation, the \(\alpha\)C-Helix residue Met290 is substantially closer to Phe382 than in the wild-type trajectory, presenting a significant steric barrier to completion of the flip.

In both the trajectories resulting from simulations of the Abl1 kinase core compound mutants, the tightly-packaged N-Lobe in the variants also puts the side chain of Ile315 closer to the positions the side chains of Asp381 or Phe382 would assume during the course of a successful flip, creating steric hindrances for either residue to flip. Beyond its flip-limiting properties as a steric obstacle, Ile315 in the constricted N-loop conformation caused by the Lys255-Glu275 or Val255-Leu273-Val280 interactions also strengthens the hydrophobic environment created by Ile293, Leu298, Leu301, Tyr353, Leu354, and Phe359. Hydrophobic as well, Phe382 is more comfortably packed in this environment in the drug-resistant variants. A comparison between the distances between Phe382 and the aforementioned hydrophobic side chains for the four representative trajectories studied is described in Supplementary Figure 9.

Additionally, both variants induce significant changes in P-Loop dynamics and conformational preference due to the interactions formed by the Lys/Val255 and B-Loop residues, which are likely to negatively impact Type-II inhibitor binding due to steric clashes and an increased barrier for the induced-fit step (which is coupled to P-Loop dynamics \cite{Ayaz2023}). This effect is compounded by the observation that C-Lobe cracking, which is also an important step for Type-II inhibitor binding \cite{Ayaz2023}, is significantly less likely in the variants (Supplementary Figures 6 and 7). 

As previously conjectured, these tight contacts and the reduced C-lobe cracking might strongly reduce the propensity for forming the DFG-out conformation, offering a common mechanism for the severe Type-II inhibitor drug resistance observed in the two compound variants: enhanced activity by destabilization of the DFG-out state, which poorly binds Type-II inhibitors. 

Since compaction of the N-Lobe by either of the Glu255Lys/Val mutations exacerbates both the steric clashes introduced by Thr315Ile and its contacts with the aforementioned hydrophobic environment, our results indicate that these variants have a clear epistatic effect in which the structural and functional consequences of each Glu255 mutation are made significantly more potent by their combination with Thr315Ile.

In prior literature, the inhibitor resistance effects of the Thr315Ile mutation were mostly attributed to steric hindrances caused by would-be inhibitors \cite{Gibbons2011}. Although studies have proposed a model in which Thr315Ile and Glu255Val/Lys act as activating mutations (shifting Abl1's conformational preference towards the active state) \cite{Yamamoto2004}, the precise mechanisms by which this shift might happen and its detailed consequences in Abl1 dynamics were not fully understood prior to the work presented above. 

Our thorough analysis of the conformational landscapes and relative state preferences of Abl1 variants bridges this gap in knowledge, providing a new level of mechanistic detail regarding the complex effects of these activating mutations. Specifically, we have found that, in addition to any ATP-binding site steric clashes with would-be inhibitors, the Glu255Lys/Val Thr315Ile mutations also significantly reduce Abl1's propensity for undergoing spontaneous inactivation through the DFG flip pathway. This shift in preference towards the active state and its neighboring conformations contextualizes the reduced affinity for Type-II inhibitors in these Abl1 variants, as Type-II inhibitors are highly specific towards the DFG-out (inactive) form of Abl1, which is much less frequently occupied in our simulations of apo Abl1 inhibitor-resistant variants.

As a crucial caveat, the conserved mechanism for inhibitor resistance we present above pertains mostly to predicted changes in inhibitor affinity during the conformational search stage of binding. Since our results show a significantly increased propensity for the inhibitor-resistant variants to adopt the active or active-adjacent states, it follows that ligands that are highly selective for the inactive state are likely to have their binding affinities reduced. However, since our simulations exclusively sampled the conformational landscape of apo Abl1, they do not offer direct insights into the impacts of the mutations on the induced-fit stage of inhibitor binding. Thus, follow-up studies exploring the impacts of mutations on the induced-fit stage of Type-II inhibitor binding on the conformational landscape of Abl1 variants in the presence of inhibitors would be crucial to complete our understanding of the effects of Abl1 drug-resistant variants on the enzyme's conformational landscape and relative state populations.

That being said, our results provide insights into the potential impacts of the mutations on the induced-fit step of binding even in the absence of direct measurements. This is because previous works have shown that specific P-Loop conformations play a significant role in the induced-fit step of Type-II inhibitor binding \cite{Wilson2015, Ayaz2023}. In our simulations, the P-Loop of wild-type Abl1 was more rigid than in the variants, which in both cases showed significant P-Loop rearrangements due to variant-only interactions between Glu/Val255 and B-Loop/\(\alpha\)c-Helix residues. The significantly increased P-Loop plasticity in inhibitor-resistant Abl1 variants is likely to affect the induced-fit step as well, potentially further reducing inhibitor affinity.

\subsection{Conservation Analysis\label{Conservation}}

To further evaluate the hypothesis that electrostatic interactions between charged residues in the N-Lobe play pivotal roles in facilitating or hindering DFG flip events in tyrosine kinases, we compared the distribution of charged residues in the Abl1 kinase core with that of other tyrosine kinases with different relative state populations than Abl1. For this comparison, we analyzed the N-Lobe charge distribution of the Src kinase core, as Src is known to occupy the active state significantly more frequently than Abl1 \cite{Wilson2015}. As a further control, we also analyzed the N-Lobe charge distribution of the Anc-AS kinase core, an ancestral kinase in-between Abl1 and Src that occupies the active state more frequently than Abl1, but not as frequently as Src \cite{Wilson2015}. The results of this comparison are summarized in Figure 10.

\begin{figure}[H]
\centering
    \includegraphics[width=500pt]{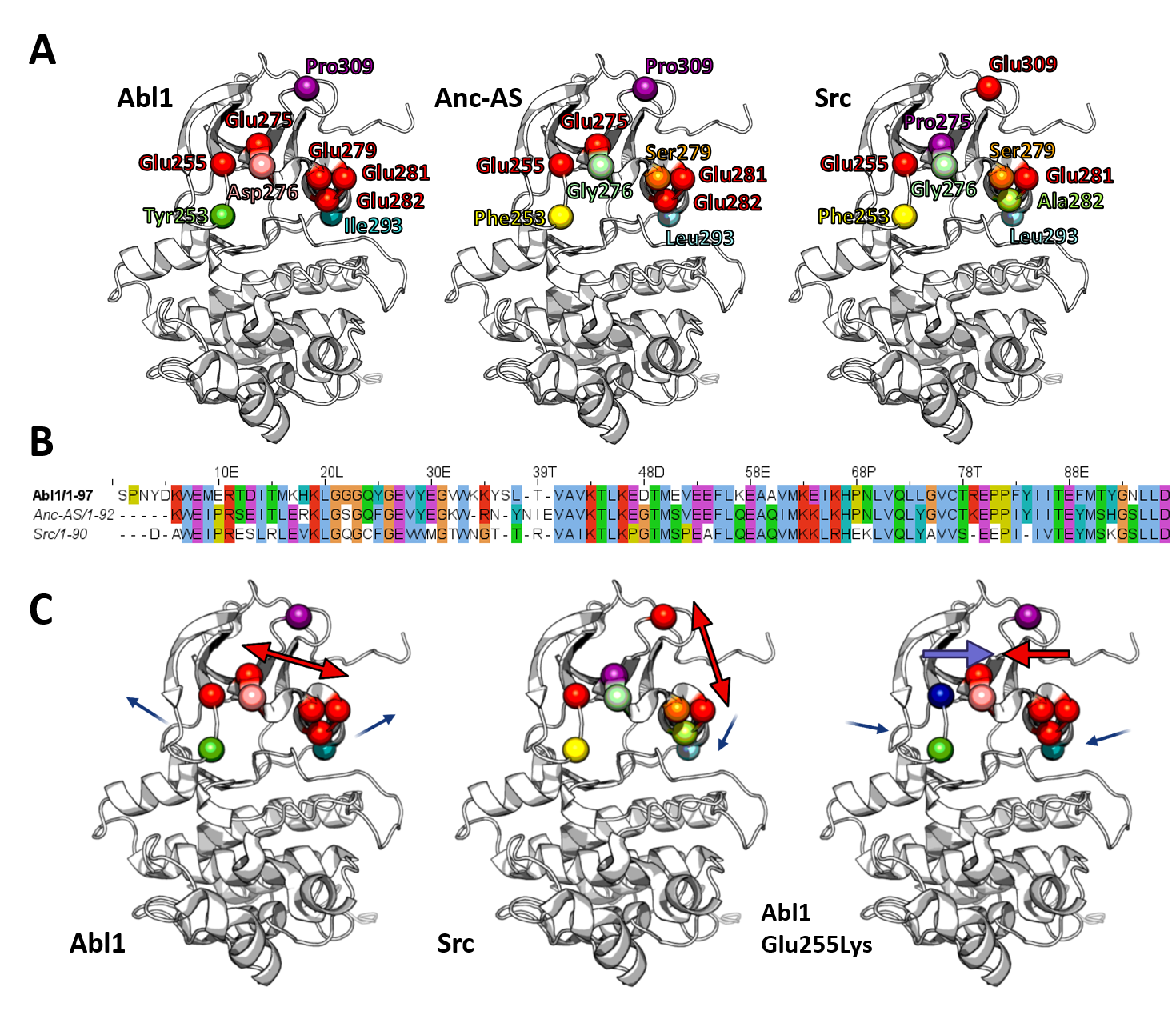}
    \caption{Conservation analysis of the charge composition of residues in the \(\alpha\)C-Helix or P-Loop of the Abl1, Ancestral AS, and Src kinases. (A) Three-dimensional structures of evolutionarily-linked kinase cores annotated with spheres marking residues within or in the close vicinity of either the P-Loop (residues 249-255 in Abl1) or \(\alpha\)C-Helix (residues 249-255) whose interactions are thought to be important for DFG flip events. (B) Alignment of the first 100 residues of the three evolutionarily-linked kinase domains with significantly different activation free energy barriers. (C) Presumed effect of different charge compositions on the backbone dynamics of the Abl1, Src, and Abl1 Glu255Lys Thr315Ile kinase cores. Colored arrows represent repulsive or attractive forces between structural elements, and thin arrows illustrate movements potentially caused by the observed charge compositions. For ease of visualization, all residue positions were annotated for the DFG-in state of wild-type Abl1 (PDB 6XR6).}
     \label{fig_10}
\end{figure}

Interestingly, the N-Lobe of Abl1 is remarkably acidic, concentrating over six negatively-charged residues (five glutamic acids and one aspartic acid) in close proximity. This charge distribution is likely to lead to strong repulsions between the triad of glutamic acids in the \(\alpha\)C-Helix (Glu279/281/282) and the negatively-charged cluster of residues near the P-Loop (Glu255/275 and Asp276). These repulsive forces could drive slight lateral expansions of the N-Lobe that would mitigate steric clashes that would prevent Asp381 or Phe382 from flipping.

In contrast with that of Abl1, the N-Lobe of Src is significantly less charged. In Src, two of the glutamic acids present in the Abl1 \(\alpha\)C-Helix triad are replaced by serine and alanine (Glu279 and Glu282, respectively), and Glu275 is replaced by a proline. This reduced charge density explains Src's increased preference for the DFG-in state compared to that of the more acidic Abl1 N-Lobe, as lateral expansion of the N-Lobe is less likely in the absence of the stronger repulsive forces present in Abl1. Finally, Abl1's proline 309 is replaced by a glutamic acid in Src. Due to this glutamic acid's close proximity to the polar and acidic residues at the start of the \(\alpha\)C-Helix, this substitution is likely to result in repulsive forces that increase the energy barrier for the \(\alpha\)C-Helix to move away from the DFG motif in Src, an effect that is absent in Abl1 (as proline is not charged), further limiting Src's ability to make space surrounding the DFG motif and thus decreasing the likelihood of the DFG flip. 

On the other hand, Anc-AS possesses an N-Lobe charge density similar to that of Abl1, with the most meaningful difference being the replacement of Glu279 in Abl1 with a serine in Anc-AS, which is also observed in Src. Since Anc-AS lies evolutionarily in-between Abl1 and Src, and previous studies have shown that its conformational preference sits roughly in-between that of both kinases \cite{Wilson2015}, the fact that the N-Lobe charge density of Anc-As also follows this pattern further points towards a link between the N-Lobe electrostatic interactions and DFG flip propensity.

Finally, both Anc-AS and Src also possess substitutions in hydrophobic residues that contact Asp381 or Phe382 at different stages of the flip: Ile293 is replaced by valine, and Tyr253 is replaced by phenylalanine. While isoleucine's side chain is significantly bulkier than valine's and thus could provide a more effective wedge for completing the Phe382 flip, the change from tyrosine to phenylalanine is harder to contextualize in this model. One hypothesis is that the side chain oxygen of Tyr253 interacts with Asp381's polar hydrogen in Abl1, orienting its side chain in a conformation that is closer to interacting with Thr315, reducing the energy barrier for the flip. This is supported by the distribution and probabilities of this observable in our WE simulations, in which Asp381 occasionally interacts with Tyr253 prior to interacting with Thr315 (Supplementary Figures 6 and 7). Further, as tyrosine is significantly more acidic than phenylalanine, the proximity of Tyr253 contributes to increasing Asp381's pKa, making it more likely to become protonated,  which significantly enhances the propensity for the flip \cite{Shan2009}. Additionally, Tyr253 mutations to phenylalanine or histidine in Abl1 are associated with Type-II inhibitor drug resistance phenotypes (Supplementary Table 1), presumably by removing the interaction with Asp381 (in the case of Tyr253Phe) or by significantly enhancing it via the resulting strong electrostatic interaction and reducing Asp381's pKa (in the case of Tyr253His).

Considering the different phenotypes of Abl1 and Src for ligand selectivity, phosphorylation rates, and conformational state preferences \cite{Wilson2015}, the observations discussed above regarding their differing N-Lobe charge densities provide strong support for the hypothesis that N-Lobe electrostatic interactions facilitate the DFG flip in Abl1 kinase, especially when corroborated by evidence from weighted ensemble simulations. Finally, mutations in the acidic residues of Abl1's N-Lobe (such as Glu255Lys and Glu255Val, the potential mechanisms for which were discussed in this study) are commonly associated with a variety of drug-resistant and oncogenic phenotypes (Supplementary Table 1), providing further credence to the presented hypothesis.

\section{Conclusions \label{conclusions}}

In this study, we present a thorough examination of the DFG flip in the Abl1 kinase core, a conformational change of significant importance for human health and drug discovery. Using enhanced sampling molecular dynamics simulations, we have obtained dozens of uncorrelated DFG flip trajectories, spanning multiple pathways and crossing an overall energy barrier of approximately 32 kcal/mol. Through our simulations, we have identified previously-unknown, potentially metastable states (DFG-inter) whose relative populations are drastically altered in inhibitor-resistant forms of Abl1, presenting a novel and comprehensive mechanism for inhibitor resistance: reduced affinity for Type-II inhibitors during the conformational search stage of ligand binding due to dramatic shifts in Abl1's propensity for undergoing spontaneous inactivation. Importantly, we show that this mechanism for Type-II inhibitor resistance is likely conserved across evolutionarily-related kinases, including Src and and Anc-AS. 

Through our work, we have simulated an ensemble of DFG flip pathways in a wild-type kinase and its variants with atomistic resolution and without the use of biasing forces. In addition, we report the atomistic impacts of inhibitor-resistant mutations on the likelihood of kinase inactivation. Since our simulations did not use biasing forces and we sampled multiple uncorrelated events, the results we have obtained for DFG flip events in Abl1 allow us to derive mechanistic insights with high confidence and unparalleled resolution. 


Our proposed mechanism synthesizes a collection of previous observations regarding the potential effects of point mutations to Glu255 and Thr315 on Abl1 dynamics and relative state populations \cite{Yamamoto2004}. Besides contributing to our general understanding of slow protein backbone dynamics in crowded micro-environments, our work creates meaningful opportunities for biomedicine, as the discovery of potentially metastable DFG states (DFG-inter) in Abl1 presents novel potential targets for the design of next-generation, pan-variant inhibitors. Additionally, the proposed general mechanism for the DFG flip can be used to orient the design of similar regulatory switches in engineered proteins. Additionally, the previously unexplored correlation we observed between the N-Lobe charge density and the propensity to undergo spontaneous conformational changes that inactivate several evolutionarily-linked tyrosine kinases could be used to contextualize interesting processes in less-understood kinases. 

Altogether, we believe that the principles revealed in this study about how proteins can experience differing dihedral dynamics as a result of varying electrostatic interactions and charge environments help to construct useful conceptual frameworks that should be of broad interest to the field of biophysics.

\section{Methods \label{Methods}}

\subsection{Molecular Dynamics and WESTPA2 Simulations\label{westpa}}

\subsection{Choice of Enhanced Sampling Methodology}

To extensively sample the active to inactive transition with maximum resolution and a broad ensemble of pathways, we chose to employ the Weighted Ensemble (WE) methodology as implemented in WESTPA2 \cite{Russo2022, Zwier2015}. In a traditional WE run, user-defined, system-dependent Progress Coordinates (PCs) are used to summarize specific motions of the system, and the chosen coordinate space is discretized into user-specified bins \cite{Zuckerman2017}. Swarms of parallel simulations (walkers) are started from one or more initial states with progress coordinates mapping to the path the users wish to sample and are allowed to evolve for a relatively short time (tau). After all walkers finish their initial evolution, the instantaneous value of each progress coordinate for each walker is evaluated, and progress is measured in terms of bin crossing events. Walkers that reach previously unoccupied bins are split (new simulations are started from the configuration assumed by the crossing walker), and walkers that remain in the same bin for a significant amount of time are merged with others in the same bin, freeing computational resources. This process is repeated for n iterations, with rigorous control of the probabilities associated with each walker so that the sum of the weights of all walkers in a WE run approximates one. 

Beyond the choice of initial states and progress coordinates, WE runs are unbiased, offering a very robust method for sampling conformational transitions in proteins. This is in contrast with enhanced sampling methods such as umbrella sampling or metadynamics in which biasing forces are applied during simulation steps and subsequently accounted for in the final measurements of properties \cite{Bussi2020}. While the aforementioned biasing methods allow us to rigorously sample processes on timescales often much longer than brute-force simulations, they often do so through significant trade-offs such as loss of information about kinetic pathways or about the existence and structure of short-lived transition states \cite{Bussi2020}. These trade-offs are absent in the traditional WE methodology (although implementations of WE with biasing forces have been used) \cite{Ray2020}, making it an excellent method for the exploration of the conformational landscape of kinases and their rare-event transitions.

\subsection{Molecular Dynamics Simulations}

Molecular dynamics simulations of the wild-type and mutant Abl1 systems were conducted using the OpenMM software package \cite{Eastman2017} with the Charmm36m force field \cite{Huang2016} and the TIP3P water model \cite{Mark2001} at 310 K and 1 ATM. For all simulations, the coordinates of active Abl1 were taken from PDB 6XR6. Systems were prepared by removing non-protein atoms, modeling relevant variants, and protonating aspartic acid 381, in accordance with previous observations that Asp381 is often protonated at physiological pHs \cite{Shan2009}.

We solvated each system within a cubic box and neutralized charges by replacing randomly selected solvent atoms with chloride and sodium ions. Following this, we used a steepest-descent algorithm to minimize the energy of each system for 50,000 minimization steps, or until the maximum force on any given atom was less than 1000 kJ/mol/min. We then equilibrated solvent atoms for 1 ns in the NVT ensemble and then for 1 ns in the NPT ensemble with solute heavy atoms restrained using the LINCS algorithm with a spring constant of 1,000 kJ/mol/m$^2$ \cite{Hess1997}. Solute heavy atom restraints were used in both the NVT and NPT equilibrations. Finally, production MD was conducted sans restraints for 50 ps per weighted ensemble iteration. For equilibration, we ran simulations with a 1 fs time step during the equilibration phase and a 2 fs time step during the production phase.

\subsection{Visualization and Analysis\label{vis}}
To visualize the predicted structures and trajectories and calculate descriptors such as distances between atoms, RMSD to reference, dihedral angles, etc., we used PyMol 2.4.1 \cite{pymol}, VMD 1.9.4 \cite{HUMP96}, MDAnalysis 2.0.0 \cite{Gowers2016, MichaudAgrawal2011}, and \cite{McGibbon2015MDTraj} 1.9.8. Data plotting and visualization were performed using Seaborn (version 0.11.1). Data were plotted as mean ± confidence intervals.

\subsection{Molecular Modeling\label{pymol}}
We used the ``Solution Builder'' module in the CHARMM-GUI web server \cite{Jo2008, Lee2015, Brooks2009} to model the drug-resistant Abl1 variants Glu255Lys Thr315Ile and Glu255Val Thr315Ile, using PDB 6XR6 (DFG-in) as the base model.

\section{Acknowledgements \label{ack}}
The authors thank Jacob Rosenstein, Joseph Larkin, F. Marty Ytreberg, Jagdish S. Patel, Marcelo D. Polêto, Winston Y. Li, and Haibo Li for insightful conversations and support.  G.M.d.S. and B.M.R. were supported in part by the National Science Foundation under Grants No. 2027108 and 2046744. G.M.d.S. has also benefited from the gracious support of the Blavatnik Family Foundation. The computational aspects of this research were conducted using computational resources and services at the Center for Computation and Visualization, Brown University.

\section{Author Contributions}
G.M.d.S. and K.L. performed the simulations and analyzed the results. D.D. and B.M.R. provided direction and oversight. G.M.d.S. and B.M.R. drafted the manuscript. All authors provided notes and edits to the manuscript.

\section{Competing Interests}
There are no competing interests. 

\section{Data and Code Availability}
The datasets from this study are available from the authors upon request. The code and scripts that underlie the simulations discussed in this study may be found in our manuscript GitHub Repository \cite{abl1_dfgflip_westpa_rawdata}.


\section{References}
\printbibliography[heading=none]


\end{document}